\documentclass[
 preprint,
superscriptaddress,
prl,
nopreprintnumbers,
 amsmath,amssymb,
 aps,
 floatfix
]{revtex4-2}

\usepackage{graphicx}
\usepackage[caption=false]{subfig}
\usepackage{color}
\usepackage{xr}
\newcommand*\Laplace{\nabla ^2}
\begin{document}

\title{Spontaneous flow instabilities of active polar fluids in three dimensions}

\author{Abhinav Singh}
 \email{absingh@mpi-cbg.de}
\affiliation{Faculty of Computer Science, Technische Universit\"{a}t Dresden, Dresden, Germany.}
\affiliation{Max Planck Institute of Molecular Cell Biology and Genetics, Dresden, Germany. }
\affiliation{Center for Systems Biology Dresden, Dresden, Germany.}
\author{Quentin Vagne}
\affiliation{University of Geneva, Geneva, Switzerland}
\author{Frank J\"{u}licher}
\affiliation{Max Planck Institute for the Physics of Complex Systems, Dresden, Germany.}
\affiliation{Center for Systems Biology Dresden, Dresden, Germany.}
\affiliation{Cluster of Excellence Physics of Life, TU Dresden, Dresden, Germany.}
\author{Ivo F. Sbalzarini}%
 \email{ivos@mpi-cbg.de}
\affiliation{Technische Universit\"{a}t Dresden, Faculty of Computer Science, Dresden, Germany.}
\affiliation{Max Planck Institute of Molecular Cell Biology and Genetics, Dresden, Germany. }
\affiliation{Center for Systems Biology Dresden, Dresden, Germany.}
\affiliation{Cluster of Excellence Physics of Life, TU Dresden, Dresden, Germany.}

\makeatletter
\newcommand*{\rom}[1]{\expandafter\@slowromancap\romannumeral #1@}
\makeatother

\begin{abstract}
Active polar fluids exhibit spontaneous flow when sufficient active stress is generated by internal molecular mechanisms. This is also referred to as an active Fr\'{e}edericksz transition. 
Experiments have revealed the existence of competing in-plane and out-of-plane instabilities in three-dimensional active matter. So far, however, a theoretical model reconciling all observations is missing. In particular, the role of boundary conditions in these instabilities still needs to be explained. Here, we characterize the spontaneous flow transition in a symmetry-preserving three-dimensional active Ericksen-Leslie model, showing that the boundary conditions select the emergent behavior.
Using nonlinear numerical solutions and linear perturbation analysis, we explain the mechanism for both in-plane and out-of-plane instabilities under extensile active stress for perpendicular polarity anchoring at the boundary, whereas parallel anchoring only permits in-plane flows under contractile stress or out-of-plane wrinkling under extensile stress.  
\end{abstract}
\maketitle

Active fluids are out-of-equilibrium materials driven by energy injection at the microscopic scale \cite{julicher_hydrodynamic_2018,bowick_symmetry_2021}. 
Active materials can have a polar or nematic alignment symmetry of the orientation vector field, and the constituents of the material allow it to generate contractile or extensile active stress. Prominent examples of active fluids are found in living matter across scales, from the cytoskeleton  \cite{kruse_generic_2005,julicher_active_2007} and tissues \cite{martin_pulsation_2010,shankar_topological_2020,guillamat_integer_2022} to collective behavior in flocks \cite{marchetti_hydrodynamics_2013}. The hydrodynamic theory of incompressible active polar fluids describes the dynamics of such active liquid crystals at long wavelengths. A key behavior of active polar fluids is their ability to generate spontaneous flow under confinement and sufficient active stress. It has been shown in two dimensions (2D) that spontaneous flow can emerge due to a Fr\'{e}edericksz-type transition first observed in passive liquid crystals \cite{freedericksz_forces_1933}. The passive Fr\'{e}edericksz transition describes the change of a homogeneous nematic state to an inhomogeneous state under the influence of external electric or magnetic fields \cite{shoarinejad_numerical_2008}. 
In active liquid crystals, the transition is driven by active molecular processes causing spontaneous material flow \cite{duclos_spontaneous_2018, ramaswamy_hybrid_2015,ramaswamy_active_2019,goldstien2012,goldstien2013,voituriez_spontaneous_2005}.

\begin{figure}
 	\centering
 	\setlength{\tabcolsep}{1pt}
 	\captionsetup[subfloat]{captionskip=-1pt} 
 	\subfloat[]{\includegraphics[scale=0.35]{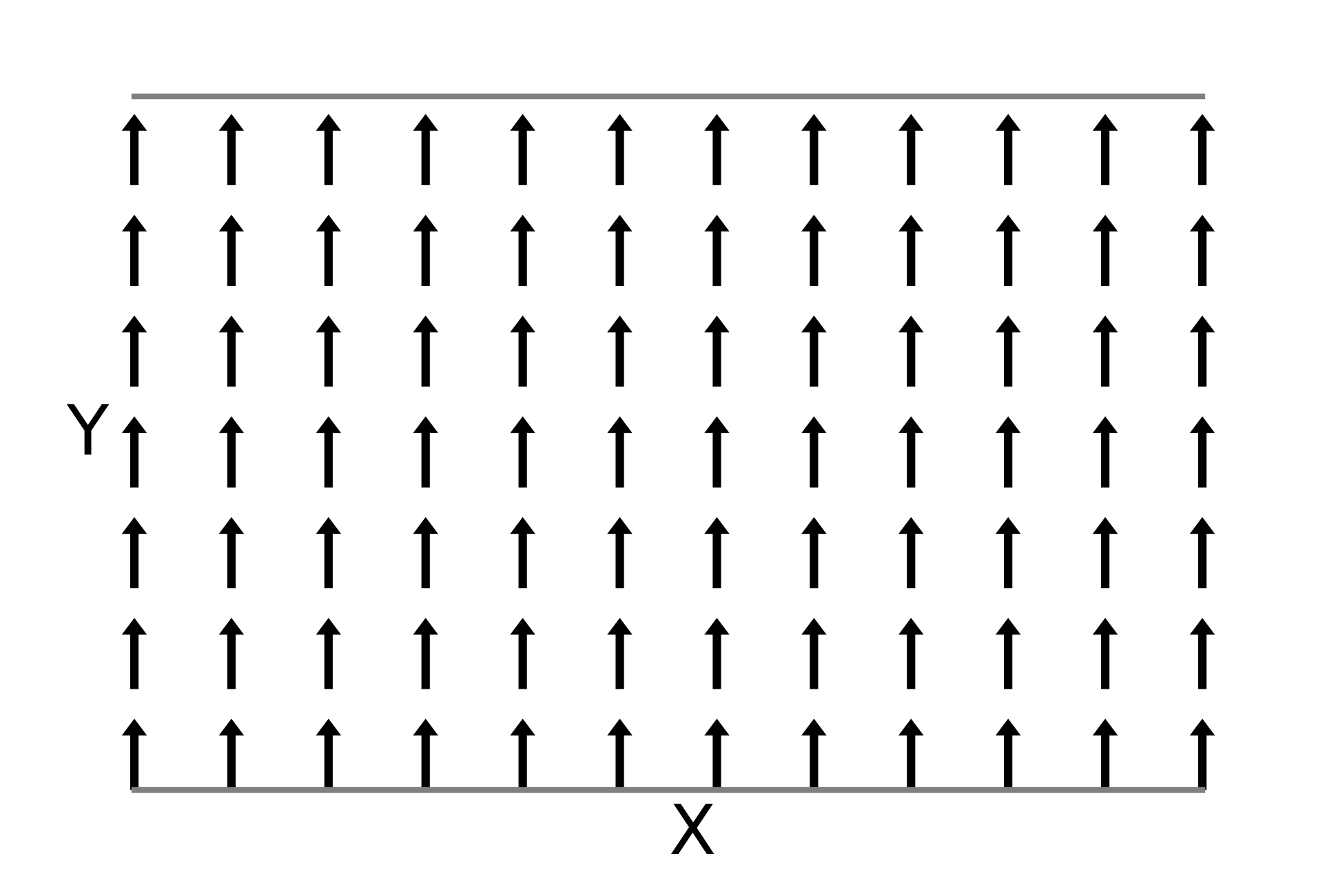}} 	\subfloat[]{ \includegraphics[scale=0.35]{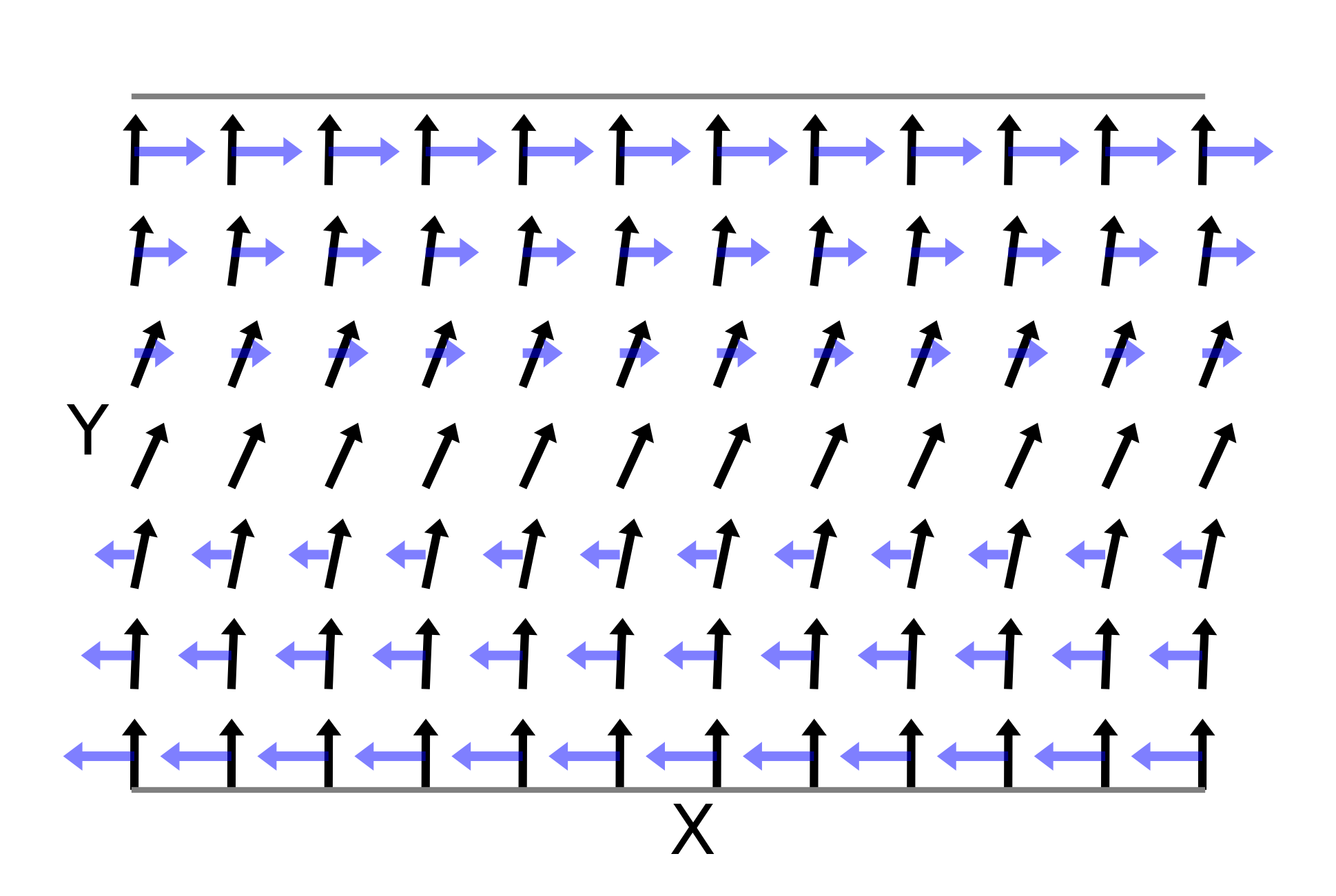}} \\[-3ex]
 	\subfloat[]{\includegraphics[scale=0.35]{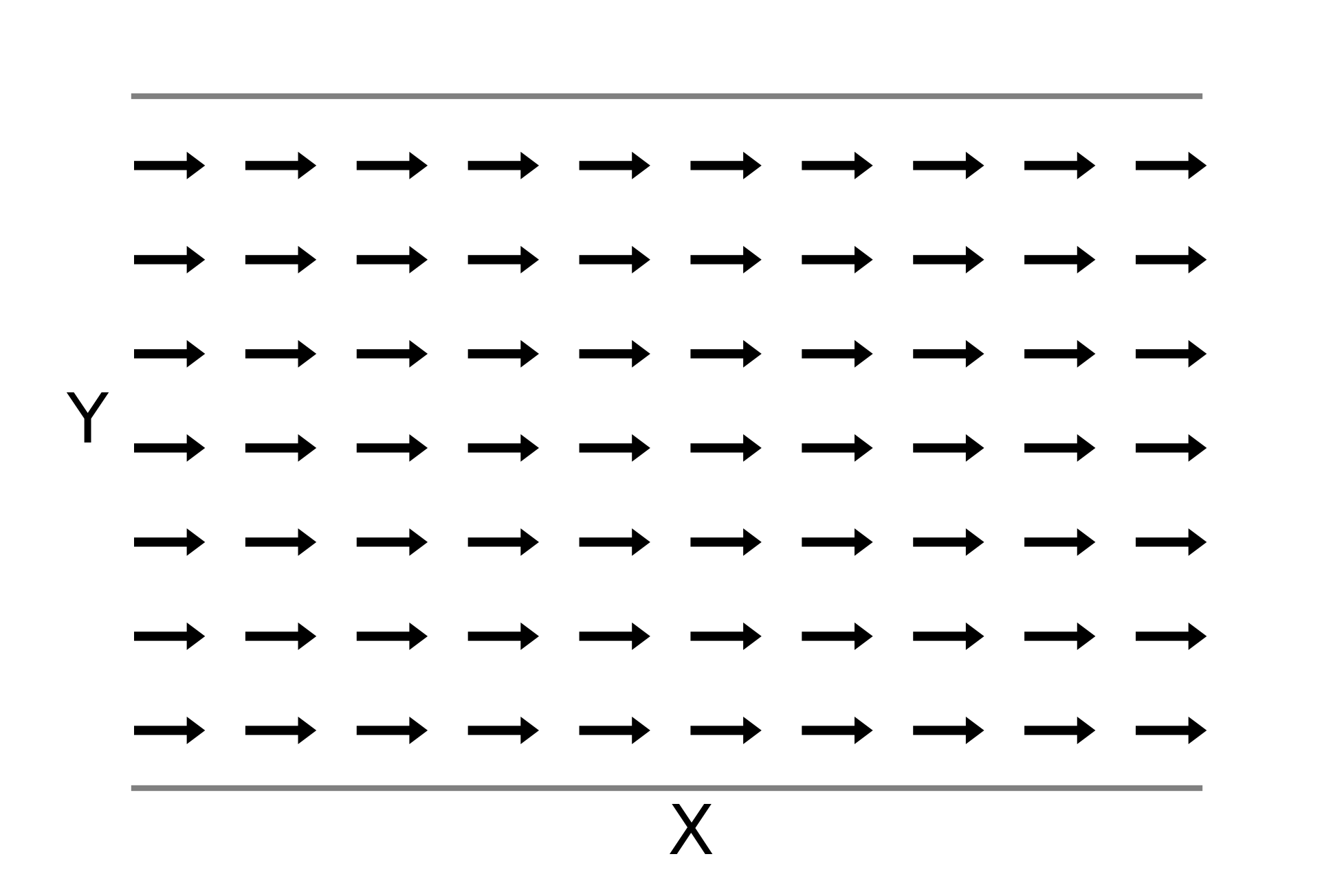}} 	\subfloat[]{\includegraphics[scale=0.35]{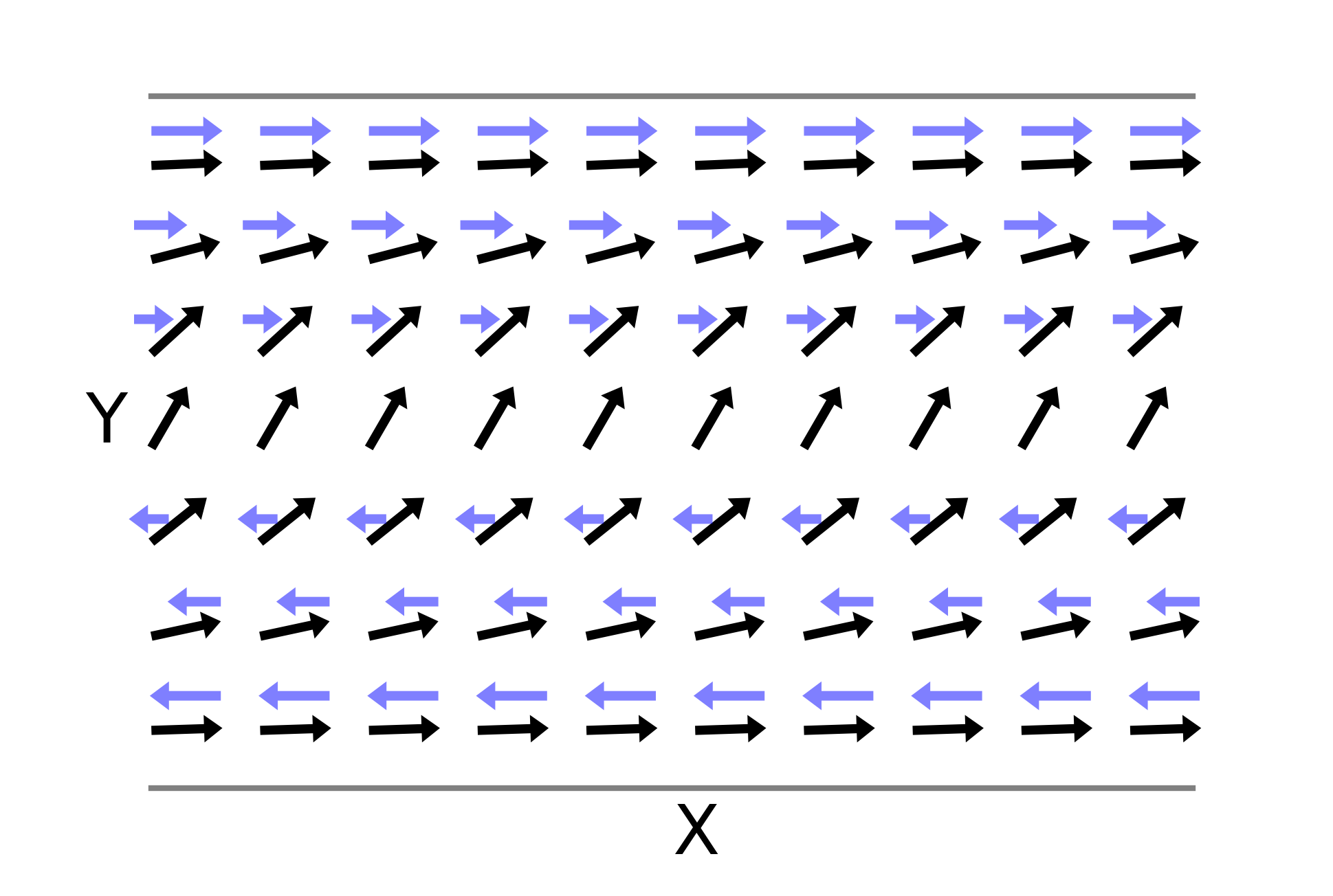}}
 	\vspace*{-0.35cm}       
     \caption{Two-dimensional active spontaneous flow transition in a plane that is infinite along X and of finite width $L$ along Y. \textbf{(a)} An initially homogeneous polarity field with perpendicular anchoring to the walls and no flow. \textbf{(b)} Spontaneous flow transition under {\em extensile} active stress with velocity indicated by blue arrows. \textbf{(c)} An initially homogeneous polarity field with parallel anchoring at the walls and no flow. \textbf{(d)} Transition to spontaneous flow under {\em contractile} active stress.}
 	\label{fig:Fred2d}       
 \end{figure}

In two dimensions, the type of instability depends on the confining boundary condition for the polarity field \cite{julicher_active_2007},  as illustrated in Fig.~\ref{fig:Fred2d}. For polarity anchored perpendicular to boundaries, one obtains a spontaneous flow transition with extensile active stress (Fig.~\ref{fig:Fred2d}a,b)~\cite{ramaswamy_hybrid_2015}. For polarity parallel to the boundary, the transition  occurs for contractile active stress (Fig.~\ref{fig:Fred2d}c,d)~\cite{voituriez_spontaneous_2005,ramaswamy_active_2019}.

Recent works suggest such instabilities to also exist in three-dimensional (3D) active polar fluids \cite{chandrakar_confinement_2020,strubing_wrinkling_2020,najma_competing_2022,Sarfati2022}. For example, an extensile active fluid was found to exhibit a bending instability in a minimal model \cite{varghese2020}, and flow-aligning active fluids were found to display coherent motion in 3D channels upon increased activity \cite{chandragiri2020}. 
Furthermore, It has been shown that 3D contractile active fluids dampen out-of-plane perturbations, whereas extensile fluids amplify them in the absence of boundary effects \cite{nejad2020}.
3D active fluids under confinement behave fundamentally different from their 2D counterparts. Notably, they exhibit flow due to buckling under extensile active stress, which is not possible in 2D for rigid boundaries \cite{chandrakar_confinement_2020}. Experimentally, such instabilities have been observed in microtubule assays capable of generating extensile active stresses \cite{chandrakar_confinement_2020,strubing_wrinkling_2020}. Some experiments suggested that in-plane and out-of-plane instabilities compete, depending on the material properties and the magnitude of the active stress~\cite{najma_competing_2022,Sarfati2022}. Other experiments~\cite{strubing_wrinkling_2020} found only an out-of-plane instability. Currently, a theoretical model explaining and reconciling all observations is missing, and the dependence of the instability on the relevant system parameters and boundary conditions remains unexplained.

Here, we study the symmetry-preserving active Ericksen-Leslie hydrodynamic model with a Lagrange multiplier enforcing constant polarity magnitude. This allows us to show that the effect of orientational order is sufficient to account for all observed instabilities in 3D.
We consider a thick active polar film, which is the 3D extension of a \textit{Fr\'{e}edericksz cell}, with anchoring of the polarity and stress-free boundary conditions on the walls.
We explain how steady-state spontaneous flows can arise for different system sizes, polarity boundary conditions, and active stress signs/magnitudes. We derive analytical expressions for the critical activity or length scale by linear perturbation analysis, analogous to the seminal work of Voituriez et al.~in 2D \cite{voituriez_spontaneous_2005}. We confirm the analytical results in convergence-validated direct 3D numerical solutions of the full nonlinear model with Lagrange multipliers and complex boundary conditions. 

\captionsetup{belowskip=-5pt}
 
\begin{figure*}
	\centering
	\setlength{\tabcolsep}{1pt}
	\begin{tabular}{ccc} 
		\subfloat[]{\includegraphics[scale=0.15]{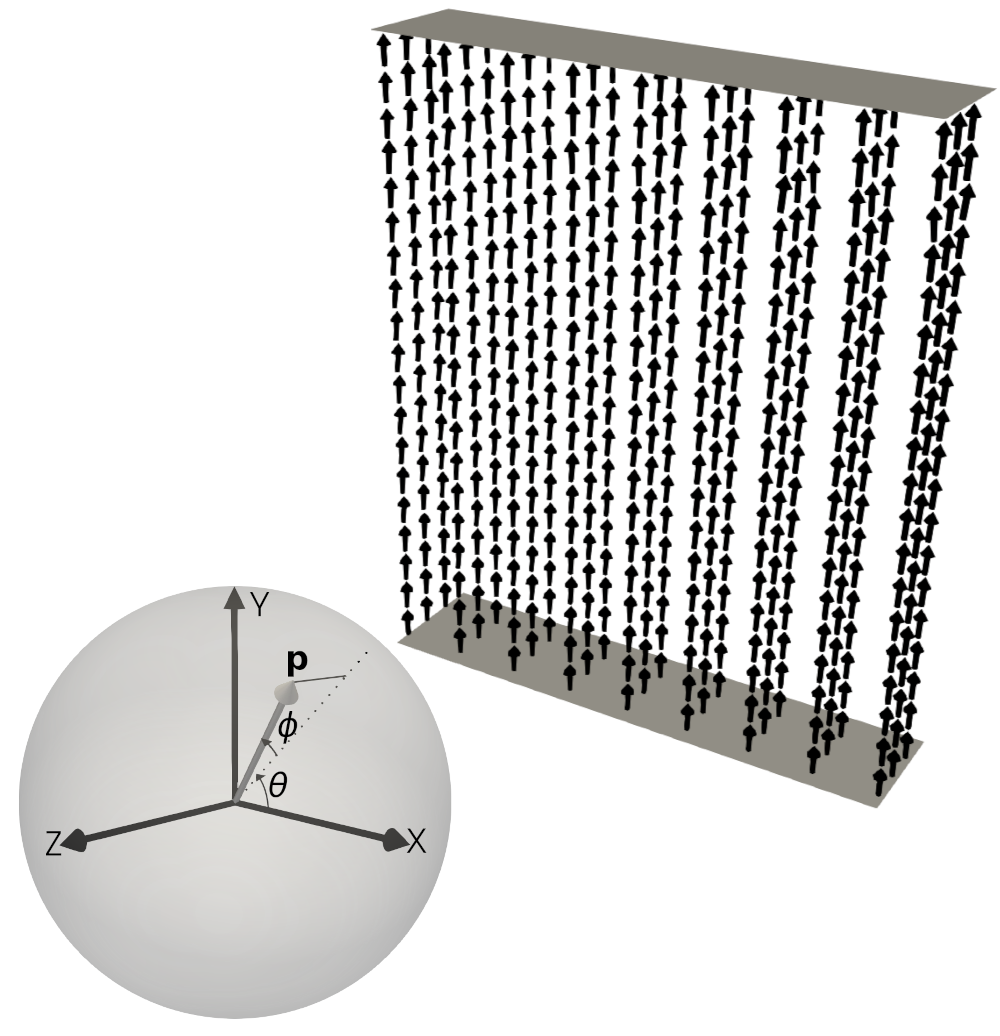}}  		&  		\subfloat[]{\includegraphics[scale=0.16]{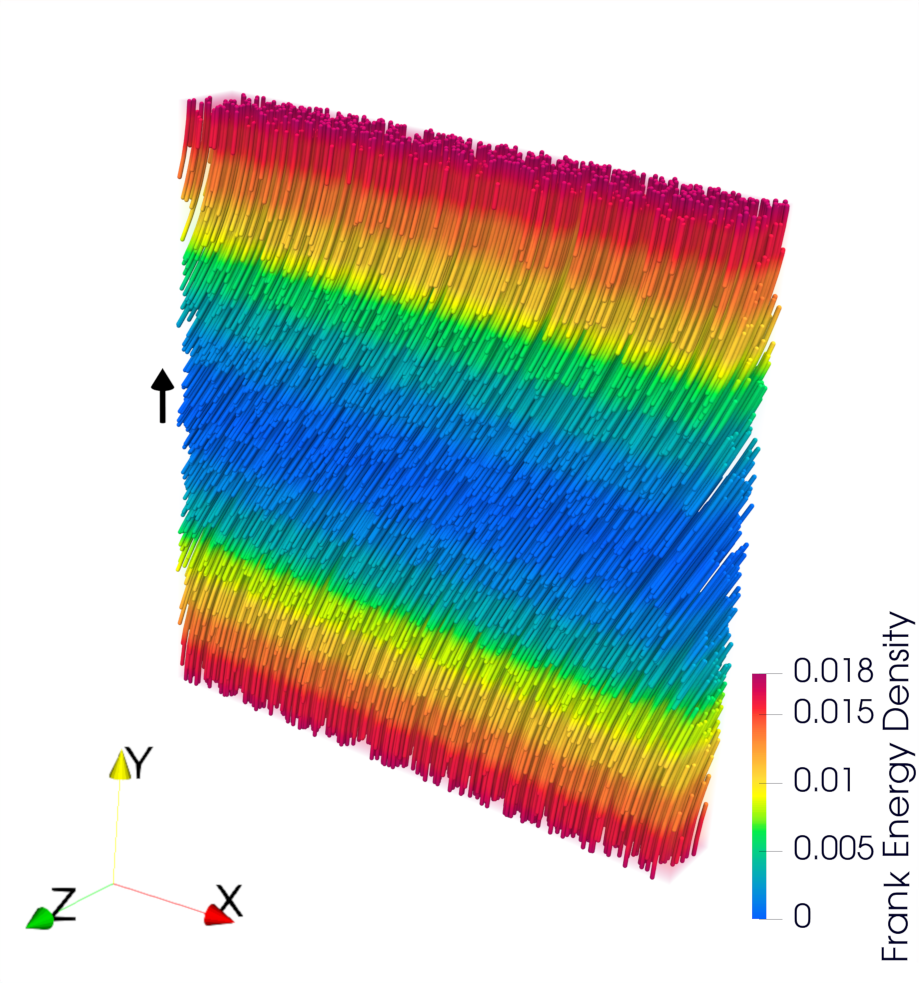}}  & \subfloat[]{\includegraphics[scale=0.16]{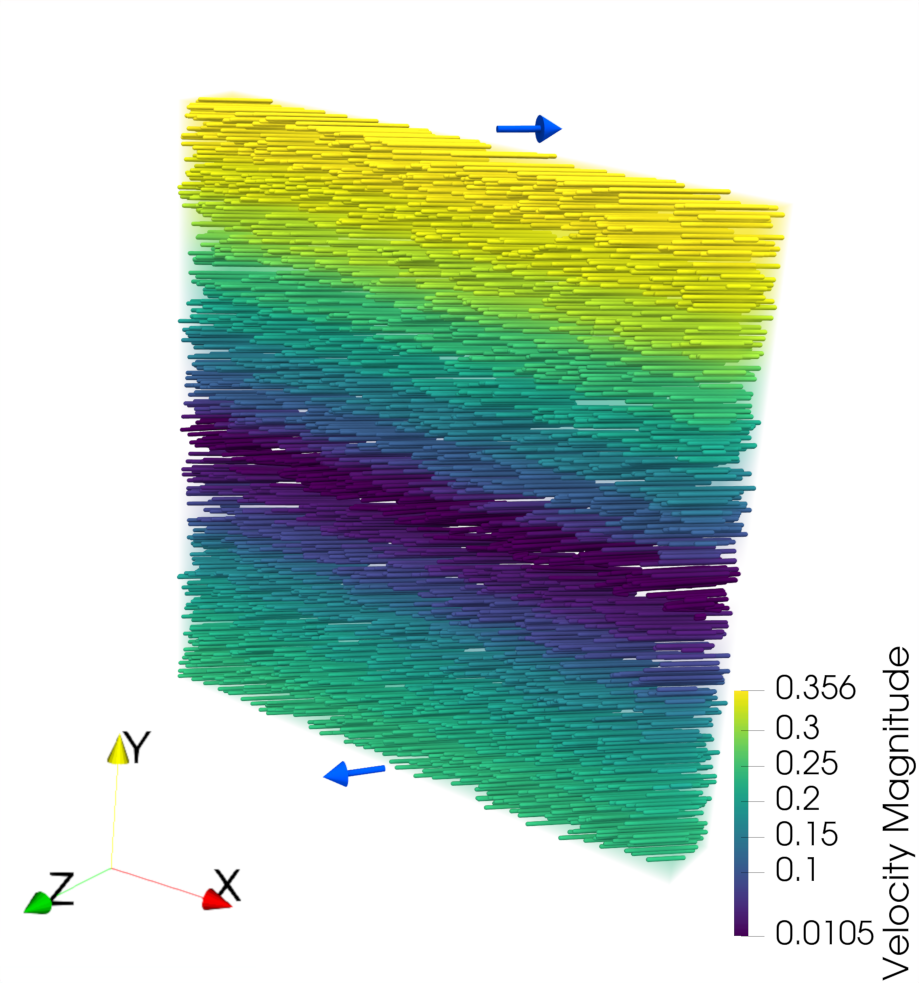}}
		\\
	\end{tabular}
	\vspace*{0.01cm}     
	\caption{Visualization of the 3D spontaneous flow transition for perpendicular polarity anchoring at the wall under extensile active stress. \textbf{(a) Coordinate system and the homogeneous steady state:} Polarity vectors below the critical active potential $(\Delta\tilde{\mu}<\Delta\tilde{\mu}_{c})$. \textbf{(b) Polarity field in the spontaneous-flow steady state:} Polarity streamlines with nonzero polarity in both $X$ and $Z$ directions, and Frank free energy density as color. The black arrow indicates the initial polarity vector direction. \textbf{(c) Velocity field in the spontaneous-flow steady state:} Velocity streamlines for the spontaneous-flow steady state in (b) with non-zero flow in both $X$ and $Z$ directions, with velocity magnitude as color. The blue arrows indicate the directions of the flow at the stress-free walls. 
	} 
	\label{fig:Freed3Dperp}    
\end{figure*}

\begin{figure*}
	\subfloat[]{\includegraphics[scale=0.14]{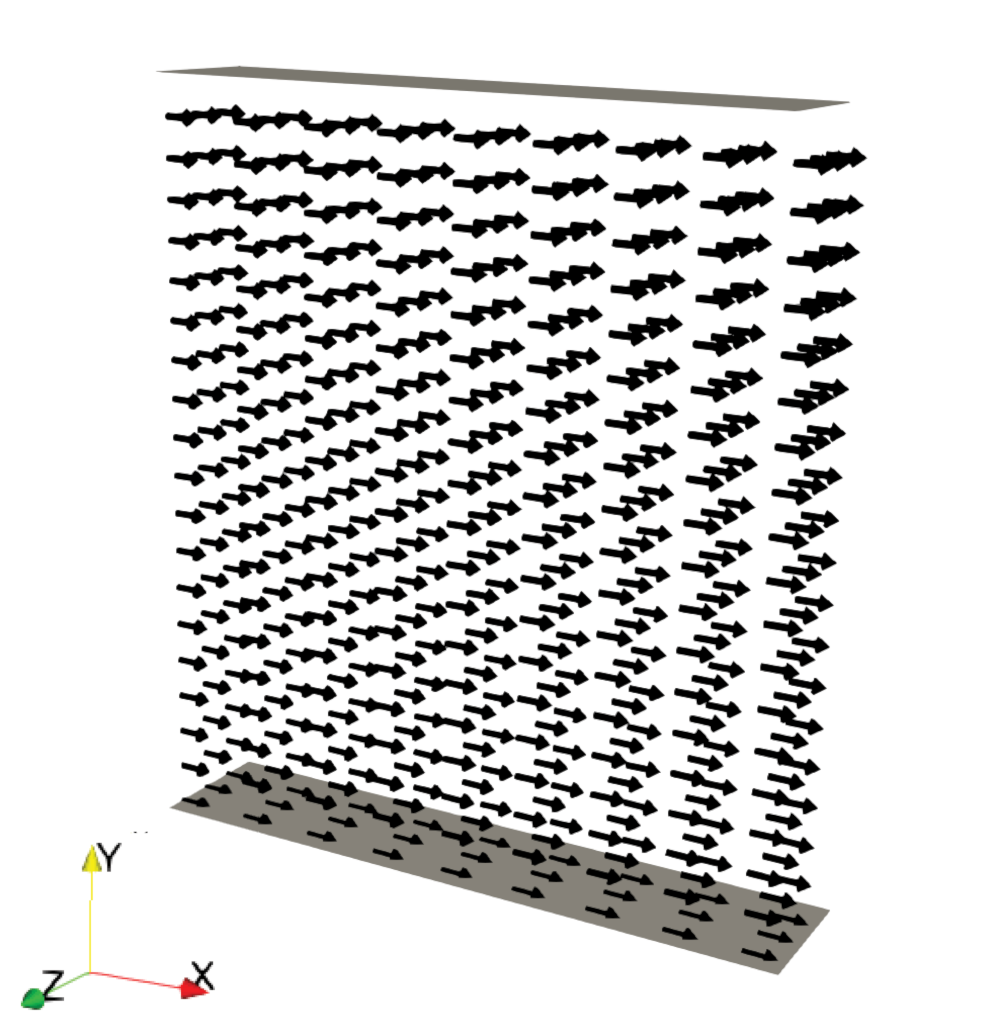}}
	\subfloat[]{\includegraphics[scale=0.15]{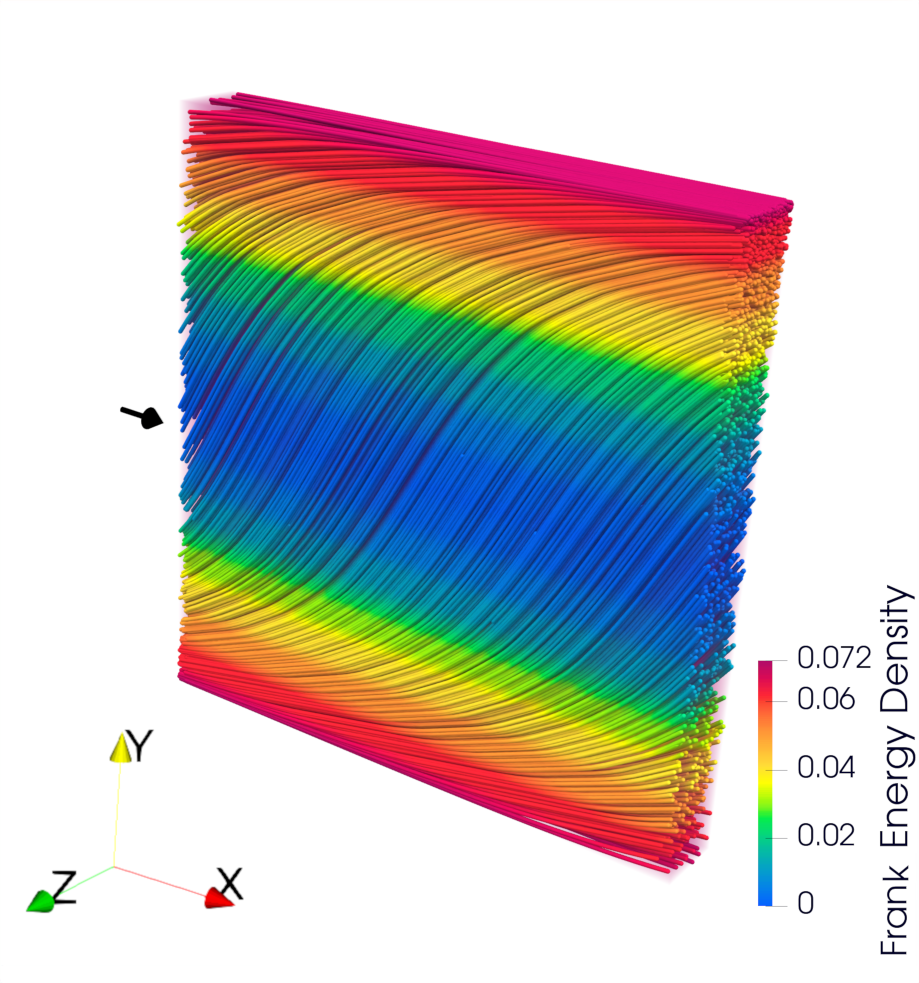} }
	\subfloat[]{\includegraphics[scale=0.15]{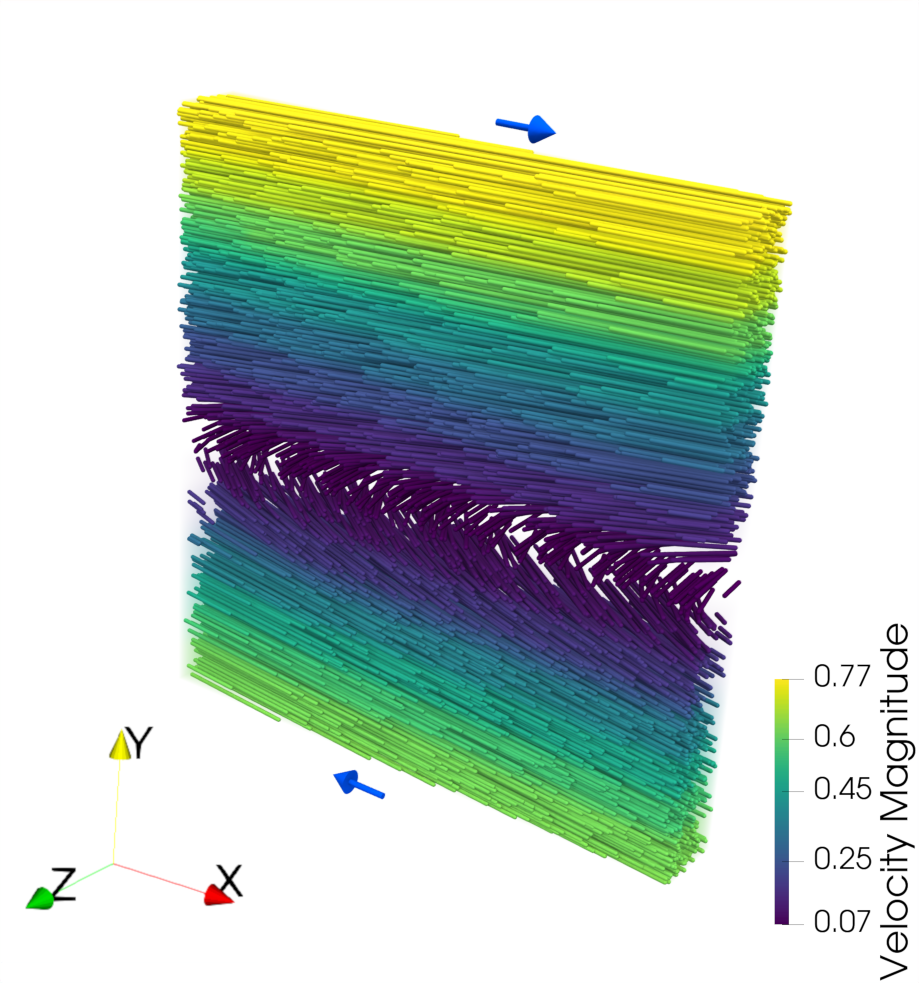}}
	\\[-3.0ex]
	\subfloat[]{\includegraphics[scale=0.15,trim=0 0 0 100,clip]{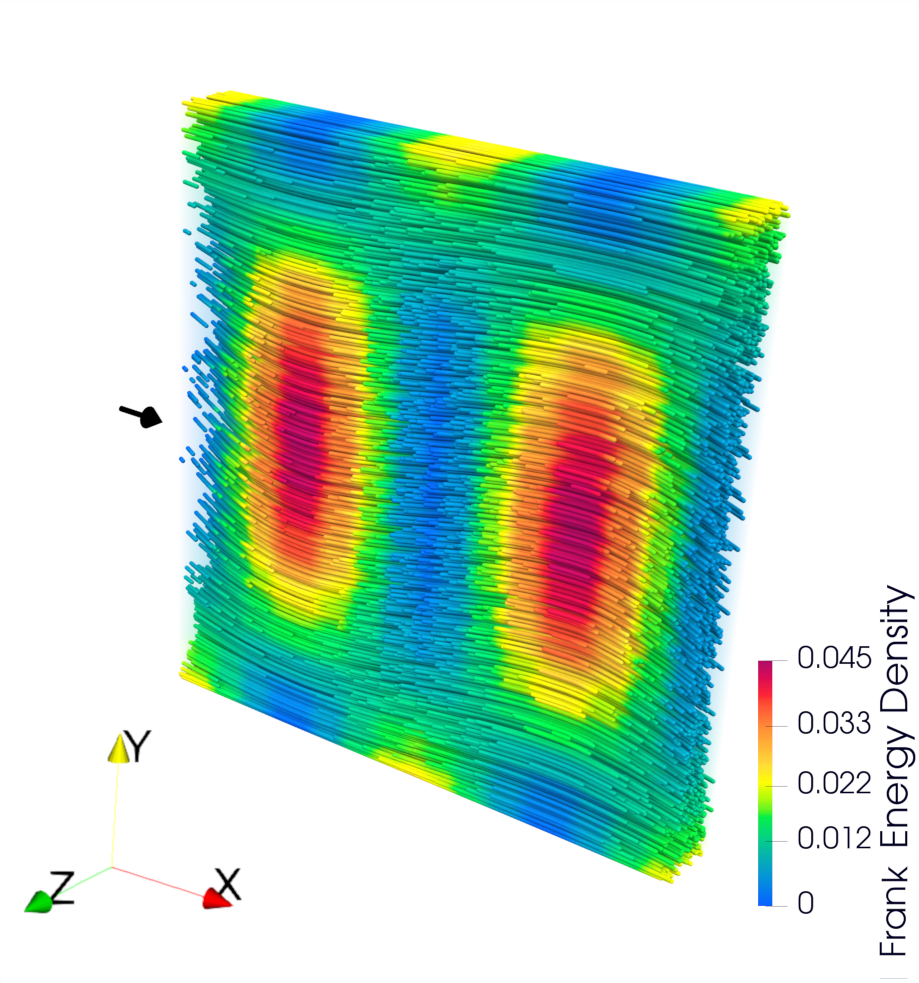}}  		 
	\subfloat[]{\includegraphics[scale=0.15,trim=0 0 0 100,clip]{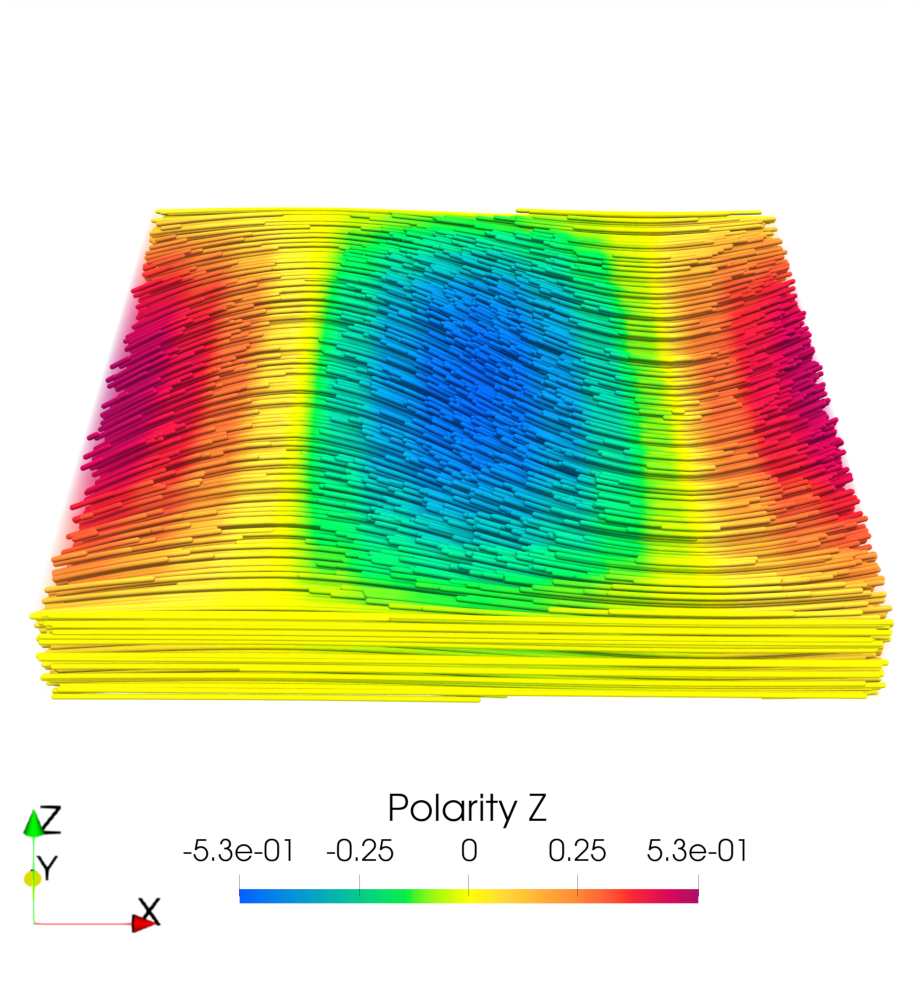}}  
	\subfloat[]{\includegraphics[scale=0.15,trim=0 0 0 90,clip]{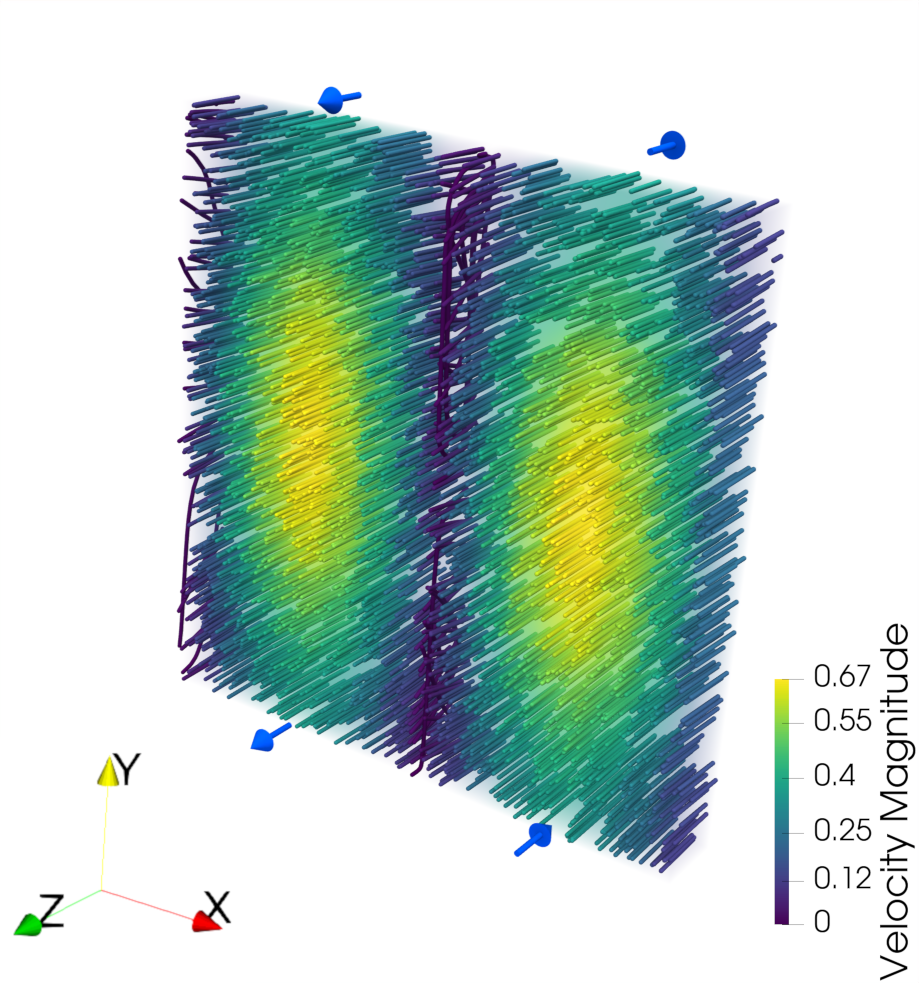}}  
	\vspace*{0.01cm}     
	\caption{Visualization of the 3D spontaneous flow transitions with parallel polarity anchoring at the wall. \textbf{(a) Homogeneous steady state:} Polarity vectors below the critical active potential ($\Delta\tilde{\mu}<\Delta\tilde{\mu}_{c}$). \textbf{(b) In-plane spontaneous-flow steady state under contractile active stress:} Polarity streamlines showing non-zero polarity in the $Y$ direction and zero in the $Z$ direction, with Frank free energy density as color. The black arrow indicates the initial polarity vector direction. \textbf{(c) Velocity field in the spontaneous-flow steady state under contractile active stress:} Velocity streamlines for the in-plane spontaneous-flow steady state in (b) with non-zero flow in the $X$ direction and velocity magnitude as color. The blue arrows indicate the directions of the flow at the stress-free walls. Parameter values for (b-c) are: $\tilde{\gamma}=1$, $\tilde{L}=10$, $\tilde{\nu}$=-0.4 , $\tilde{\zeta}=-1$, $\Delta\tilde{\mu}=0.35$. \textbf{(d) Out-of-plane spontaneous-flow steady state under extensile active stress:} Polarity streamlines showing non-zero polarity field in the $Z$ direction and zero in the $Y$ direction, with Frank free energy density as color. The black arrow indicates the initial polarity vector direction. \textbf{(e) $Z$ component of the polarity field in the out-of-plane spontaneous-flow steady state under extensile active stress:} \textbf{(f) Velocity in the out-of-plane spontaneous-flow steady state under extensile active stress:} Velocity streamlines for the spontaneous-flow steady state under extensile active stress in (d) with nonzero flow in the $Z$ direction and velocity magnitude as color. The blue arrows indicate the directions of the flow at the stress-free walls. Parameter values for (d-f) are: $\tilde{\gamma}=1$, $\tilde{L}=10$, $\tilde{\nu}$=-0.4 , $\tilde{\zeta}=1$ $\Delta\tilde{\mu}=2.4$.}
	\label{fig:Freed3Dpar}       
\end{figure*}

We find a transition in 3D for perpendicular anchoring  under extensile active stress (Fig.~\ref{fig:Freed3Dperp}). This transition is different from the 2D case (Fig.~\ref{fig:Fred2d}a) as the resulting shear flow is both along the $X$ and $Z$ directions (Fig.~\ref{fig:Freed3Dperp}c). This leads to out-of-plane bending of the polarity and to a 3D spontaneous flow transition that is invariantly extended along the $X$ and $Z$ directions (Fig.~\ref{fig:Freed3Dperp}b). For parallel anchoring at the wall, we find a transition with contractile active stress that impedes out-of-plane perturbations and is an invariant extension of the 2D spontaneous flow transition (Fig.~\ref{fig:Freed3Dpar}b-c), and a purely out-of-plane ``wrinkling'' under extensile active stress that does not exist in 2D (Fig.~\ref{fig:Freed3Dpar}d-f). 

{ \bf{Hydrodynamics of active polar fluids.}} 
The incompressible viscous active polar fluid equations \cite{julicher_hydrodynamic_2018} can be described in Einstein summation notation as:
\begin{subequations}
	\begin{eqnarray}
		\frac{\mathrm{D} p_{\alpha}}{\mathrm{D} t}=\frac{h_{\alpha}}{\gamma}-\nu u_{\alpha \beta} p_{\beta}+\lambda\Delta\mu p_\alpha \label{eqP1}\\
		\partial_{\beta} \sigma_{\alpha \beta}^{(\text{tot})}-\partial_{\alpha} \Pi=0 \label{eqP2},~
		\partial_{\gamma} v_{\gamma}=0\label{eqP3}\\
		2 \eta u_{\alpha \beta}=\sigma_{\alpha \beta}^{(s)}+\zeta \Delta \mu\left(p_{\alpha} p_{\beta}-\frac{1}{3} p_{\gamma} p_{\gamma} \delta_{\alpha \beta}\right)\nonumber\\
		-\frac{\nu}{2}\left(p_{\alpha} h_{\beta}+p_{\beta} h_{\alpha}-\frac{2}{3} p_{\gamma} h_{\gamma} \delta_{\alpha \beta}\right), \label{eqP4}
	\end{eqnarray}
	\label{eq::activeGel}
\end{subequations}
with $\alpha,\beta,\gamma \in \{x,y,z\}$ for the spatial components. Further details can be found in the Supplementary Information. This model fulfills the Onsager symmetry relations and accounts for stresses from elastic distortion of the nematic field, as well as antisymmetric and Ericksen stresses. The scalar $\nu$ is the standard liquid-crystal flow aligning/tumbling parameter.
The molecular field $\mathbf{h}=K\Laplace \mathbf{p}+h_{\Vert}^0\mathbf{p}$ can be decomposed into parallel $h_\Vert=\mathbf{p}\cdot\mathbf{h}$ and perpendicular $\mathbf{h_\perp}=\mathbf{p}\times\mathbf{h}$ components in a local co-moving frame. Note that molecular fields differing by a factor of $h_\Vert^0\mathbf{p}$ are equivalent~\cite{DeGennes1995} and hence $h_\Vert=h_\Vert^0$. Using Eq.~(\ref{eqP1}) and enforcing $p_\gamma\frac{Dp_\gamma}{Dt}=0$, $h_\Vert$ can be derived as $
	h_\Vert=-\gamma\Big[\lambda \Delta \mu-\frac{2\nu}{p_\gamma p_\gamma}\Big(
	u_{\alpha\beta}p_\alpha p_\beta\Big)\Big],	\label{eq:lag}
$
such that $\|\mathbf{p}\|$ remains constant.
In 2D, the perpendicular component of the molecular field is a scalar with one degree of freedom, which makes the nonlinear equations analytically tractable. In 3D, however, $\mathbf{h_\perp}=(h_{\perp x},h_{\perp y},h_{\perp z})$ is a 3-vector, and the coupling of the Frank free energy with the Lagrange multiplier $h_{||}$ renders the force-balance equation intricate.
Therefore, it is not clear from simplified models how the flow instability depends on system parameters such as $\lambda$, $\zeta$, or $\nu$~\cite{varghese2020,chandrakar_confinement_2020}. The microscopic origin of $\lambda$ and $\zeta$, however, 
can be elucidated in the full nonlinear model, providing experimentally measurable predictions.
\begin{figure*}
	\centering
	\subfloat[]{\includegraphics[scale=0.32,trim=30 11 10 0,clip]{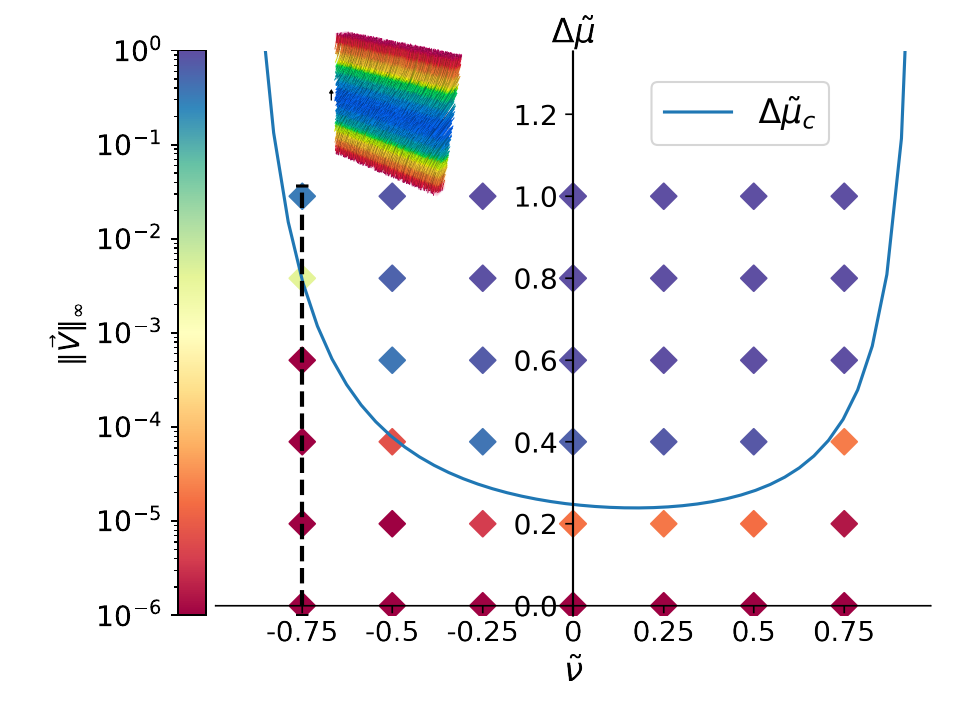}} 	\subfloat[]{\includegraphics[scale=0.32,trim=30 11 20 0,clip]{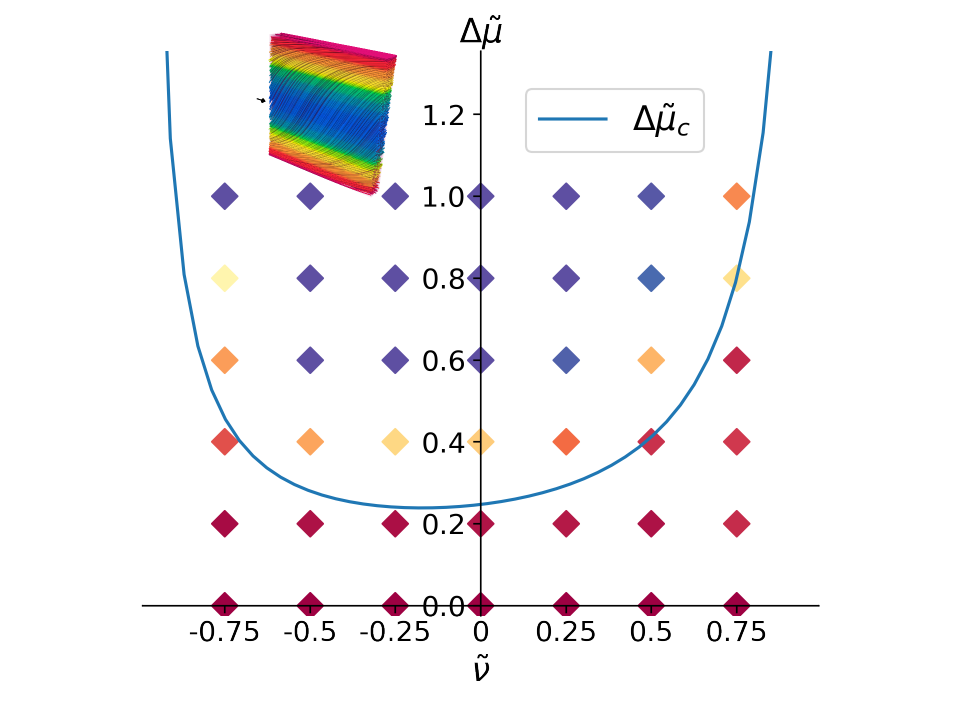}}    \subfloat[]{\includegraphics[scale=0.32,trim=85 11 40 0,clip]{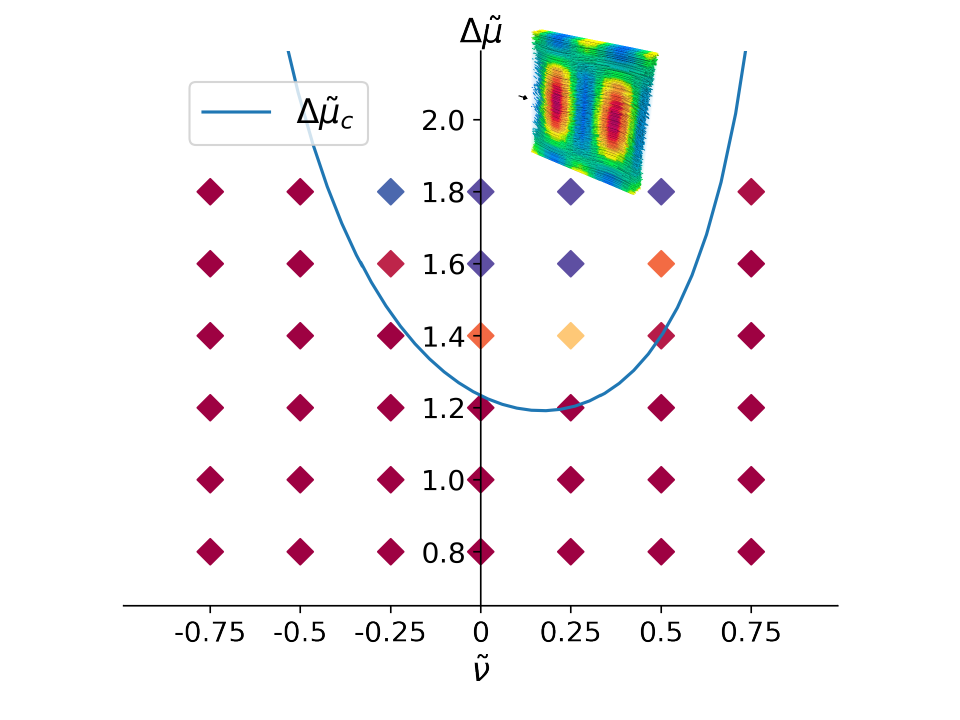}}  
	\subfloat[]{\includegraphics[scale=0.32,trim=60 0 0 0,clip]{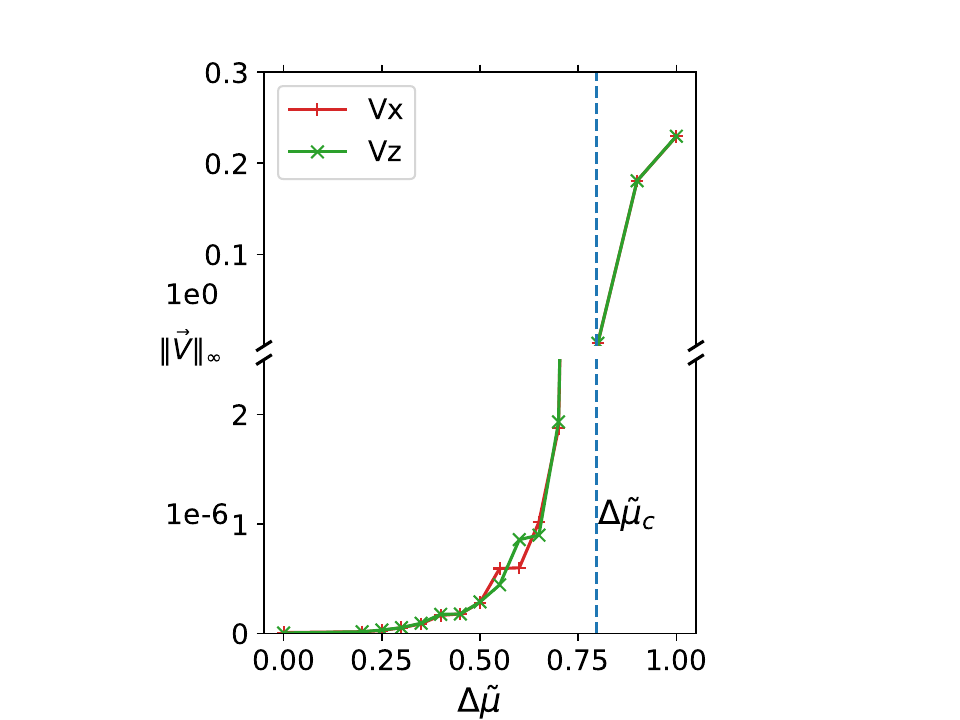}	\label{fig:exprise}} 
	\vspace*{0.01cm}     
	\caption{Transition for different flow regimes. The solid lines plot the analytical expressions for the dimensionless critical activity $\Delta\tilde{\mu}_c$ vs.~dimensionless flow-tumbling parameter $\tilde{\nu}$ for $\tilde{\gamma}=\tilde{\zeta}=1$ and  $\tilde{L}=10$. The color of the $\blacklozenge$ symbols in the background grid indicates the maximum norm of the spontaneous flow velocity obtained by numerically solving the nonlinear equations at those parameters with a tolerance of $10^{-6}$.
		\textbf{(a)} Perpendicular polarity anchoring with extensile stress, Eq.~(\ref{eq:mu1b}). See (d) for more simulations along the dashed line. \textbf{(b)} Parallel anchoring with contractile stress, Eq.~(\ref{eq:mu2b}); \textbf{(c)} parallel anchoring with extensile stress, Eq.~(\ref{eq:mu3b});
		\textbf{(d)} Numerically obtained maximum flow velocity magnitude vs.~activity $\Delta\tilde{\mu}$ for $\tilde{\nu}=-0.75$ (dashed vertical line in (a)) for spontaneous flow with perpendicular anchoring and extensile active stress. Note the broken Y-axis with different scales to accommodate for the sharp increase around the critical activity $\Delta\tilde{\mu}_{c}$.}
	\label{fig:nu3D}     
\end{figure*}
{ \bf{Results.}}
We analyze the hydrodynamic equations (\ref{eq::activeGel}) at steady state and provide a mechanism for the emergence of  symmetry-breaking spontaneous flow.
We express the unit polarity vector as $\mathbf{p}=[\cos(\theta)\cos(\phi),\, \sin(\theta)\cos(\phi),\, \sin(\phi)]$
 using the coordinates illustrated in figure~\ref{fig:Freed3Dperp}. We eliminate the velocity from Eq. \eqref{eq::activeGel} and obtain the nonlinear force balance equation as a function of $(\theta,\phi)$. For anchoring boundary conditions $(\theta_0,\phi_0)$ at the top and bottom walls, the system is steady with trivial solution $[\theta(y),\phi(y)]=(\theta_0,\phi_0)$. 
 
For the case of perpendicular anchoring of the polarity at the boundary, i.e., $(\theta_0,\phi_0)=(\frac{\pi}{2},0)$. 
Assuming small perturbations $[\epsilon(y), \kappa(y)]$ around the no flow homogeneous steady state, we obtain the equations for the perturbations by linearizing the nonlinear equation:
\begin{equation}
	K\frac{\partial^2}{\partial y^2}\begin{bmatrix}  \epsilon(y) \\ \kappa(y)  \end{bmatrix}
	=\frac{2 \gamma  \Delta\mu  (\nu -1)   (\zeta +\gamma  \lambda  \nu )}{\gamma  (\nu -1)^2+4 \eta }
	\begin{bmatrix}\epsilon(y)\\ \kappa(y) \end{bmatrix}.
	\label{eq:p1}
\end{equation}
Hence, there are perturbation modes of the form
\begin{equation}
\begin{bmatrix}
\epsilon(y) \\
\kappa(y)
\end{bmatrix}		=\sin \left(\frac{\pi}{L} y\right)
\begin{bmatrix}
\epsilon_{\mathrm{m}} \\
\kappa_{\mathrm{m}} 
\end{bmatrix},
			\label{eq:m1b}
\end{equation}
where $\epsilon_{\mathrm{m}},\kappa_{\mathrm{m}}$ are the maximum tilt magnitudes in $\theta$ and $\phi$, respectively. These modes effectively define a critical activity
\begin{equation}
		\Delta\mu_{c}=\frac{K\pi ^2 \left(4 \eta+\gamma  (\nu -1)^2 \right)}{2L^2  \gamma  (\nu -1) (-\zeta -\gamma  \lambda  \nu )}\, .
\label{eq:mu1b}
\end{equation}
Assuming $-1<\nu<1$, i.e., a flow-tumbling active fluid, we note that for the critical $\Delta\mu_c>0$ the effective active stress is extensile, $-\zeta-\gamma\lambda\nu<0$. 
Hence for a flow-tumbling polarity, we expect the spontaneous flow transition when  $\Delta\mu>\Delta\mu_c$. However, the instability depends on a nonlinear combination of parameters. For example, the critical activity depends nonlinearly on $\nu$. Further, the effects of $\zeta$ and $\lambda$ are coupled to both the rotational viscosity $\gamma$ and to $\nu$. For $|\nu|>1$, the behavior depends on $|\zeta|$. We provide phase diagrams for these three qualitatively different critical behaviors in Supplementary Information SI-3.

The governing equations can be nondimensionalized with respect to $\lambda$, $\eta$, and $K$ by rescaling $\tilde{L}=(\eta\lambda)^{\frac{1}{3}}L$, $\tilde{\zeta}=\zeta (\eta\lambda)^{-1}$, $\Delta\tilde{\mu}=\Delta\mu (\eta\lambda)^{\frac{1}{3}}K^{-1}$, $\tilde{\gamma}=\gamma\eta^{-1}$, and $\tilde{\nu}=\nu$. We numerically solve the dimensionless equations for a thick 3D active film that is periodic along the $X$ and the $Z$ directions and has thickness $L$ in the Y direction. Details of the simulation method can be found in SI-2. The simulation computer code scales to parallel computer architectures, as it is based on the open-source scientific computing library OpenFPM~\cite{incardona_openfpm_2019} and a template expression language for partial differential equations \cite{singh_c_2021}.

We verify the expressions in Eq.~(\ref{eq:mu1b}) by numerically solving the nonlinear equations for $\tilde{\gamma}=1$, $\tilde{\zeta}=1$, $\tilde{L}=10$. For these parameters, the dependence of the critical activity $(\Delta\tilde{\mu}_{c})$ on the flow-tumbling parameter $\tilde{\nu}$ is shown in figure~\ref{fig:nu3D}a as a solid line. Hence, when $\Delta\tilde{\mu}>\Delta\tilde{\mu}_c$,  
the mode in Eq.~(\ref{eq:m1b}) appears with spontaneous flow governed by
\begin{equation}
	\frac{\partial}{\partial y}\begin{bmatrix}v_x(y) \\ v_z(y)
	\end{bmatrix}=\frac{2 \Delta\mu  (\gamma  \lambda  \nu +\zeta )}{\gamma  (\nu -1)^2+4 \eta }
	\begin{bmatrix}
		-\epsilon (y) \\
		\kappa(y)
	\end{bmatrix}.
	\label{eq:mu1v}
\end{equation}

We find that the mode may be stabilized by the nonlinearities above the critical potentials, resulting in steady-state flows. In the steady flow state, active stresses are balanced by elastic nematic stresses, leading to stationary perturbation modes.
The corresponding spontaneous flow transition is observed above the critical activity and is visualized in figure \ref{fig:Freed3Dperp} and SI-Video \rom{1}. 

We study the stability of parallel anchoring of the polarity at the boundary, i.e., $(\theta_0,\phi_0)=(0,0)$. We again linearize around small angle perturbations  $(\epsilon,\kappa)$ of the polarity and obtain: 
\begin{equation}
		K\frac{\partial^2}{\partial y^2}\begin{bmatrix}\epsilon(y)\\ 
\kappa(y)
		\end{bmatrix}
	=\frac{2 \gamma \Delta \mu  (\nu +1) (\gamma  \lambda  \nu +\zeta )}{\gamma  (\nu +1)^2+4 \eta }
	\begin{bmatrix}
\epsilon(y)\\0
	\end{bmatrix}
		.
\end{equation}
This leads to a critical activity
\begin{equation}
\Delta\mu_{c}=\frac{\pi ^2 K \left(4 \eta+\gamma  (\nu +1)^2 \right)}{2 \gamma  L^2 (\nu +1) (-\zeta-\gamma  \lambda  \nu )}\, .
\label{eq:mu2b}
\end{equation}
For the critical $\Delta\mu_c>0$, the active stress for the transition is contractile, $-\zeta-\gamma\lambda\nu>0$ giving rise to a spontaneous flow transition with an S-like shape of the polarity field (figure \ref{fig:Freed3Dpar}b), as observed in 2D, but invariantly extended in the third dimension. This confirms that contractile active polar fluids impede out-of-plane perturbations when polarity is anchored on the boundary. 
The spontaneous flow transition occurs when $\Delta\mu>\Delta\mu_{c}$.
We confirm this in nonlinear simulations for the same parameters as before and $\tilde{\zeta}=-1$.
The dependence of the critical activity for contractile active stress on the flow-tumbling parameter $\tilde{\nu}$ is plotted in figure~\ref{fig:nu3D}b as a solid line. The numerical solutions for $\Delta\tilde{\mu}>\Delta\tilde{\mu}_{c}$ confirm the transition for different values of $\nu$, the maximum norm of the flow velocity is shown as a color code in figure~\ref{fig:nu3D}b. The initial homogeneous state of polarity is shown in figure \ref{fig:Freed3Dpar}a. Figure~\ref{fig:Freed3Dpar}b shows the polarity field for the steady-state spontaneous flow shown in figure \ref{fig:Freed3Dpar}c. The onset of this transition is also shown in SI-Video \rom{2}.

We note that the symmetry can also be broken in the $X$ direction instead of $Y$, such that $u_{xz}\neq 0$ and $u_{xy}= 0$. A similar analysis then reveals a 2D extensile perturbation mode:
\begin{equation}
K\Laplace_{\{x,y\}}\begin{bmatrix}
\epsilon(x,y)\\
\kappa(x,y) 
\end{bmatrix}	=
\frac{-2 \gamma  \Delta \mu  (\nu -1) (\gamma  \lambda  \nu +\zeta )}{\gamma  \left(\nu-1\right)^2+4 \eta }
	\begin{bmatrix}
		0\\
	\kappa(x,y)
	\end{bmatrix}.
\end{equation}
This leads to perturbations of the form 
\begin{equation}
\kappa(x,y)=\kappa_m\cos\left(\frac{2 \pi}{L_x}x+\alpha\right)\sin\left(\frac{\pi}{L_y}y\right),
\label{eq:2dmode}
\end{equation}
where $\alpha$ is the phase shift, which is fixed by the flow boundary condition and the integration constant. This mode corresponds to a critical activity
\begin{equation}
	\Delta\mu _{c}=\frac{(4L_x^2+L_y^2) \pi ^2 K \left(\gamma  \left(\nu -1\right)^2+4 \eta \right)}{2 \gamma  L_x^2L_y^2 (\nu -1) (-\zeta -\gamma  \lambda  \nu)}\, .
	\label{eq:mu3b}
\end{equation}
For the critical $\Delta\mu_c>0$, the active stress for the transition is extensile, $-\zeta-\gamma\lambda\nu<0$.
The mode in Eq.~(\ref{eq:2dmode}) describes an out-of-plane transition maintaining $\theta(x,y)=0$. For the previously chosen parameters and $\tilde{\zeta}=1$, the dependence of $\Delta\tilde{\mu}_{c}$ on $\tilde{\nu}$ is shown in figure~\ref{fig:nu3D}c.

The above expressions further clarify the effect of the finite length $L_x$ of the domain in the X direction. In the ideal physical system, $L_x$ is infinite, and $L_y$ is finite. In Eq.~(\ref{eq:mu3b}), we see that there is a non-zero limit for the critical activity as $L_x$ approaches infinity, and the mode of deformation in Eq.~\eqref{eq:2dmode} has no modulation in the X direction. This predicts the wrinkling wavelength close to the transition. For further increasing activity, however, the perturbations are no longer small, rendering the linearized equations invalid. Then, the system transitions to spatiotemporal chaos.

The amplitudes $\epsilon_{\mathrm{m}},\kappa_{\mathrm{m}}$ depend on $\Delta\tilde{\mu}$ and are analytically intractable. The critical active potential for out-of-plane wrinkling is significantly larger than for the other cases. This causes the instability to occur earlier or faster in time in 3D. The unstable mode shows oscillatory flows in opposite directions. With $\Delta \tilde{\mu}>\Delta\tilde{\mu}_{c}$, we find a spontaneous flow transition of small amplitude near the critical value, as shown by the maximum norm of the velocity (color of symbols). The associated wrinkling in the transition is shown in figure \ref{fig:Freed3Dpar}d--f and SI-Video \rom{3}.
This transition has also been observed experimentally in extensile polar fluids and referred to as {\em bending} or {\em wrinkling instability} \cite{chandrakar_confinement_2020,strubing_wrinkling_2020,najma_competing_2022,Sarfati2022}.  Here, we qualitatively characterized the effect of finite channel length $L_x$ on the wrinkling wavelength.

{ \bf{Conclusions.}} 
We have derived the critical active stress for the spontaneous flow transition in 3D active liquid crystals from the full, symmetry-preserving active Ericksen-Leslie model with Lagrange multipliers. We found that contractile active stresses impede out-of-plane perturbations at the transition, whereas extensile active stresses promote them under parallel anchoring of the polarity at the walls. For perpendicular polarity boundary conditions, we found a 3D active Fr\'{e}edericksz-type transition under purely extensile stress. We analytically derived the critical active potentials for the transition in each case and confirmed them in direct numerical solutions of the nonlinear 3D system. For a fixed activity $\Delta\mu$, the present analysis equivalently yields a corresponding critical length $L_c$ that defines a system size above which the transition occurs. The results show how the instabilities arise from the interplay between the boundary conditions, the active potential, and the channel width and length. 

Our work can be related to previous studies on similar systems.
Similar in-plane and out-of-plane instabilities as described here were previously reported using hybrid lattice-Boltzmann simulations with periodic boundary conditions \cite{nejad2020}. However, for flow-tumbling active fluids, another study reported no coherent flow in 3D channels without preferential anchoring on the boundaries \cite{chandragiri2020}. Here we have shown that such flows do occur if the polarity field is anchored at the surfaces.
Interestingly, a simplified model generated pumping behavior without anchoring boundary conditions when no-slip velocity boundary conditions were used~\cite{varghese2020}. We verified that for the no-slip boundary conditions, such behavior also occurs in our model for super-critical activity. We then observe the same flow modes as previously seen in 2D \cite{voituriez_spontaneous_2005}.

Our results accurately predict the type of spontaneous flow in 3D and provide a comprehensive understanding of 3D active fluids, unifying previously made observations and explaining their physical origin. We have shown how 3D active matter differs from its 2D counterpart due to the additional degrees of freedom. Indeed, we found a bending or wrinkling instability mediating spontaneous flow in 3D, explaining earlier experimental observations~\cite{chandrakar_confinement_2020,strubing_wrinkling_2020,najma_competing_2022,Sarfati2022} with direct implications for understanding biological morphogenesis and finding design principles for the control of active matter.

\vspace{2mm}
{ \textit{Acknowledgements.}} 
This work was supported by the Center for Scalable Data Analytics and Artificial  Intelligence (ScaDS.AI) Dresden/Leipzig, funded by the Federal Ministry of Education and Research (BMBF, Bundesministerium f\"{u}r Bildung und Forschung).
We thank the Scientific Computing Facility of MPI-CBG and the Center for Information Services and High Performance Computing (ZIH) of TU Dresden for
providing the compute resources for the numerical simulations. We thank the anonymous referees for their constructive comments.

\end{document}


\maketitle
\section{3D Numerical Solution of Active Polar Hydrodynamics}\label{simdetails}
In Einstein summation notation, the incompressible viscous active polar fluid equations are 
\cite{julicher_hydrodynamic_2018}:
\begin{subequations}
	\begin{eqnarray}
		\frac{\mathrm{D} p_{\alpha}}{\mathrm{D} t}=\frac{h_{\alpha}}{\gamma}-\nu u_{\alpha \beta} p_{\beta}+\lambda\Delta\mu p_\alpha+
		\omega_{\alpha \beta} p_{\beta}\label{eqP1}\\
		\partial_{\beta} \sigma^{(\mathrm{tot})}_{\alpha \beta}-\partial_{\alpha} \Pi=0 \label{eqP2}\\
		\partial_{\gamma} v_{\gamma}=0\label{eqP3}\\
		2 \eta u_{\alpha \beta}=\sigma_{\alpha \beta}^{(s)}+\zeta \Delta \mu\left(p_{\alpha} p_{\beta}-\frac{1}{3} p_{\gamma} p_{\gamma} \delta_{\alpha \beta}\right)-\frac{\nu}{2}\left(p_{\alpha} h_{\beta}+p_{\beta} h_{\alpha}-\frac{2}{3} p_{\gamma} h_{\gamma} \delta_{\alpha \beta}\right). \label{eqP4}
	\end{eqnarray}
	\label{eq::activeGelSI}
\end{subequations}
The time evolution of the polarity field 
$\mathbf{p}=(p_\mathrm{x},p_\mathrm{y},p_\mathrm{z})^\top$  is governed by Equation~(\ref{eqP1}).
The co-rotational Lagrangian derivative is defined as $
\frac{\mathrm{D} p_{\alpha}}{\mathrm{D} t}=\frac{\partial p_{\alpha}}{\partial t}+v_{\gamma} \partial_{\gamma} p_{\alpha}+\omega_{\alpha \beta} p_{\beta},$
where $\omega_{\alpha \beta}=\frac{1}{2}\left(\partial_{\alpha} v_{\beta}-\partial_{\beta} v_{\alpha}\right)$ is the vorticity tensor.
$u_{\alpha \beta}=\frac{1}{2}\left(\partial_{\alpha} v_{\beta}+\partial_{\beta} v_{\alpha}\right)$ is the strain rate tensor, $\gamma$ is the rotational viscosity of the polarity field, $\nu$ is the coupling coefficient for mechanical stress and polarization that controls the flow-aligning ($|\nu|>1$) or flow-tumbling ($|\nu|<1$) nature of the active fluid. $\lambda$ is the coefficient coupling the polarity dynamics with the active chemical potential $\Delta\mu$. 
We decompose the molecular field $\mathbf{h}$ into parallel and perpendicular components,
\begin{subequations}
	\begin{eqnarray}
		h_\Vert = \mathbf{p}\cdot\mathbf{h} =p_xh_x+p_yh_y+p_zh_z\\
		\mathbf{h_\perp}=\mathbf{p}\times\mathbf{h} = (h_{\perp x},h_{\perp y},h_{\perp z})\nonumber\\=(p_yh_z- p_zh_y, p_zh_x-p_xh_z, p_xh_y - p_yh_x).
	\end{eqnarray}
	\label{eq:hpp}
\end{subequations}
The vector $\mathbf{h_\perp}$ is computed from the variational derivative of the Frank free energy density
\begin{equation}
	F_{3D}=\frac{K_s}{2} (\nabla\cdot \mathbf{p})^2+ \frac{K_t}{2} (\mathbf{p}\cdot \nabla\times \mathbf{p}){}^2 + \frac{K_b}{2}  (\mathbf{p}\times(\nabla \times \mathbf{p}))^2 - \frac{1}{2}h^0_{\Vert}\Vert \mathbf{p}\Vert^2
	\label{eq:freeEnergy3d}
\end{equation} with respect to $\mathbf{p}$.
The total stress $\sigma^{(\mathrm{tot})}_{\alpha\beta}=\sigma^{(\mathrm{s})}_{\alpha\beta}+\sigma_{\alpha \beta}^{(\mathrm{ant})}+\sigma^{(\mathrm{e})}_{\alpha\beta}$ is decomposed as the sum of the symmetric ($\mathrm{s}$), antisymmetric ($\mathrm{ant}$), and equilibrium ($\mathrm{e}$) stresses. The equilibrium stress, also called the Ericksen stress, is given by
\begin{equation}
	\sigma_{\alpha \beta}^{(\mathrm{e})}=-\frac{\partial F_{3D}}{\partial\left(\partial_{\beta} p_{\gamma}\right)} \partial_{\alpha} p_{\gamma},
\end{equation}
with $F_{3D}$ from Eq.~(\ref{eq:freeEnergy3d}). 
The anti-symmetric stress is
\begin{equation}
	\sigma_{\alpha \beta}^{(\mathrm{ant})}=\frac{1}{2}\left(p_{\alpha} h_{\beta}-p_{\beta} h_{\alpha}\right).
\end{equation}
Setting $p_\gamma \frac{Dp_\gamma}{Dt}=0$ to maintain constant polarity magnitude $p_\gamma p_\gamma$, we derive the Lagrange multiplier
\begin{equation}
	h_\Vert=-\gamma\Big[\lambda \Delta \mu-\frac{\nu}{p_x^2+p_y^2+p_z^2}\Big(u_{xx}p_x^2+u_{yy}p_y^2+u_{zz}p_z^2+
	2u_{xy}p_xp_y+2u_{yz}p_yp_z+2u_{xz}p_xp_z \Big)\Big].		\label{eq:lag}
\end{equation}
Substituting the decomposition of $\mathbf{h}$
\begin{subequations}
	\begin{eqnarray}
		h_x=h_\Vert p_x  - h_{\perp z}p_y + h_{\perp z}p_y\\
		h_y=h_\Vert p_y + h_{\perp z}p_x - h_{\perp x}p_z\\
		h_z=h_\Vert p_z + h_{\perp x}p_y - h_{\perp y}p_x
	\end{eqnarray}
\end{subequations}
with the Lagrange multiplier from Eq.~(\ref{eq:lag}) and combining it with the force-balance Eq.~(\ref{eqP2}), we derive the steady-state component-wise Stokes flow equations that are implemented in computer code using a custom C++ expression system \cite{singh_c_2021} in the scalable scientific computing library OpenFPM~\cite{incardona_openfpm_2019}. 
At time 0, the polarity is homogeneously aligned with the anchoring boundary condition except a point perturbation of 0.001 radians in both positive Y and Z directions at $x=L/2$ to break the symmetry.

The time evolution of the polarity is computed using Adams-Bashforth-Moulton predictor-corrector 
time integration with a time step of 0.01 and renormalization of the slopes with a final time of $t_f=100$. The steady state is detected with a tolerance of $10^{-8}$. Note that near the critical activity, the transition can be very slow and difficult to catch numerically. Hence, for the simulations shown in Fig.~(4d) of the main text, we used a more accurate direct solver for the velocity from the MUMPS library \cite{amestoy_performance_2019}, which is based on LU-decomposition, and we increased the spatial resolution from $18\times19\times5$ to $64\times65\times5$ grid points for higher accuracy. Further, a smaller absolute tolerance of $10^{-11}$ and relative tolerance of $10^{-9}$ were used for adaptive time stepping of the same stepper.

The velocity field is computed by iteratively correcting pressure and solving the implicit system of incompressible Ericksen-Leslie Stokes equations with hydrodynamic stress-free boundary conditions and the constraint of no flow at $x=L/2$, $y=L/2$. At each time step, the resulting linear system of equations is solved numerically using the iterative GMRES solver as implemented in the PETSc software library~\cite{balay_efficient_1997}. We checked that using a higher resolution yields the same results, confirming grid convergence.

\section{Derivation of Critical Activity}
Follwing \cite{voituriez_spontaneous_2005}, we derive $\mathbf{h_\perp}$  from the model equations at steady state and then equate it to the $h_\perp$  based on the Frank free energy to analyze the response to a perturbation in 3D. We consider a thick film that is infinitely extended along the X and Z directions and has a thickness of $L$ in the Y direction. The surface of the film at $y = L$ and $y=0$ is stress-free ($\sigma_{xy}  = 
0$ and $\sigma_{yz}  = 0$), 
and impenetrable $(v_y (x, y, z, t) = 0)$.  The polarity is fixed on the top ($y=L$) and bottom ($y=0$) such that $(p_x, p_y, p_z)  = (\cos (\theta_0) \cos (\phi_0) ,\, \sin (\theta_0) \cos (\phi_0) ,\,  \sin (\phi_0) )$. Under these conditions, $v_y = 0$ everywhere due to incompressibility and translation invariance in X and Z directions. Further, $~u_{xy}=\partial_yv_x,~u_{yz}=\partial_yv_z$ and $u_{xx}=u_{yy}=u_{zz}=u_{xz}=0$. 

Fixing polarity to be perpendicular to the boundary wall, i.e., $(\theta_0,\phi_0)=(\pi/2,0)$, and assuming small perturbations $\epsilon(y),\kappa(y)$, the restoring force up to linear order of tilt is $\mathbf{h_\perp}=(K\frac{\partial \kappa(y)}{\partial  y^2},0,K\frac{\partial \epsilon(y)}{\partial y^2})$, where $K=K_s=K_t=K_b$ is the elastic constant in the single-constant approximation of the Frank free energy.
Using Eq.~(\ref{eqP4}) and imposing $\sigma_{xy}^{(tot)}=\sigma_{zy}^{(tot)}=0$, we obtain the strain rates $u_{xy}, u_{yz}$ and substitute them in Eq.~(\ref{eqP1}) to obtain the force associated with $\mathbf{h_\perp}$. 
The so-obtained non-linear equation is decoupled from the flow and only depends on $(\theta,\phi)$ as shown in the supplementary Mathematica notebook. We do not reproduce this equation here due to its excessive length.
Substituting into this equation a small perturbation $(\epsilon(y),\kappa(y))$ and linearizing around $(\theta_0,\phi_0)=(\frac{\pi}{2},0)$, 
we obtain the dynamical equation of the perturbation as described in the main text.
This leads to a spontaneous flow transition under extensile active stress. 
Up to linear orders of tilt, the strain rates in this case are:
\begin{equation}
	\frac{\partial}{\partial y}\begin{bmatrix}v_x(y) \\ v_z(y)
	\end{bmatrix}=\frac{2 \Delta\mu  (\gamma  \lambda  \nu +\zeta )}{\gamma  (\nu -1)^2+4 \eta }
	\begin{bmatrix}
		-\epsilon (y) \\
		\kappa(y)
	\end{bmatrix}.
	\label{eq:mu1v}
\end{equation}

We repeat the analysis for $(\theta_0,\phi_0)=(0,0)$ and obtain the critical activity of a 3D spontaneous flow transition under contractile active stress. In this case, the restoring force up to linear order of tilt is $\mathbf{h_\perp}=(0,-K\frac{\partial \kappa(y)}{\partial  y^2},K\frac{\partial \epsilon(y)}{\partial y^2})$. 
Up to linear order of tilt, the strain rate in this case is:
\begin{equation}
	\frac{\partial v_x(y)}{\partial y}=\frac{2 \Delta\mu  (\gamma  \lambda  \nu +\zeta )}{\gamma  (\nu +1)^2+4 \eta }\epsilon(y) .
\end{equation} 

\begin{figure}
	\subfloat[Perpendicular anchoring with extensile stress. $\zeta=1,\nu=-1.5$]{\includegraphics[scale=0.25]{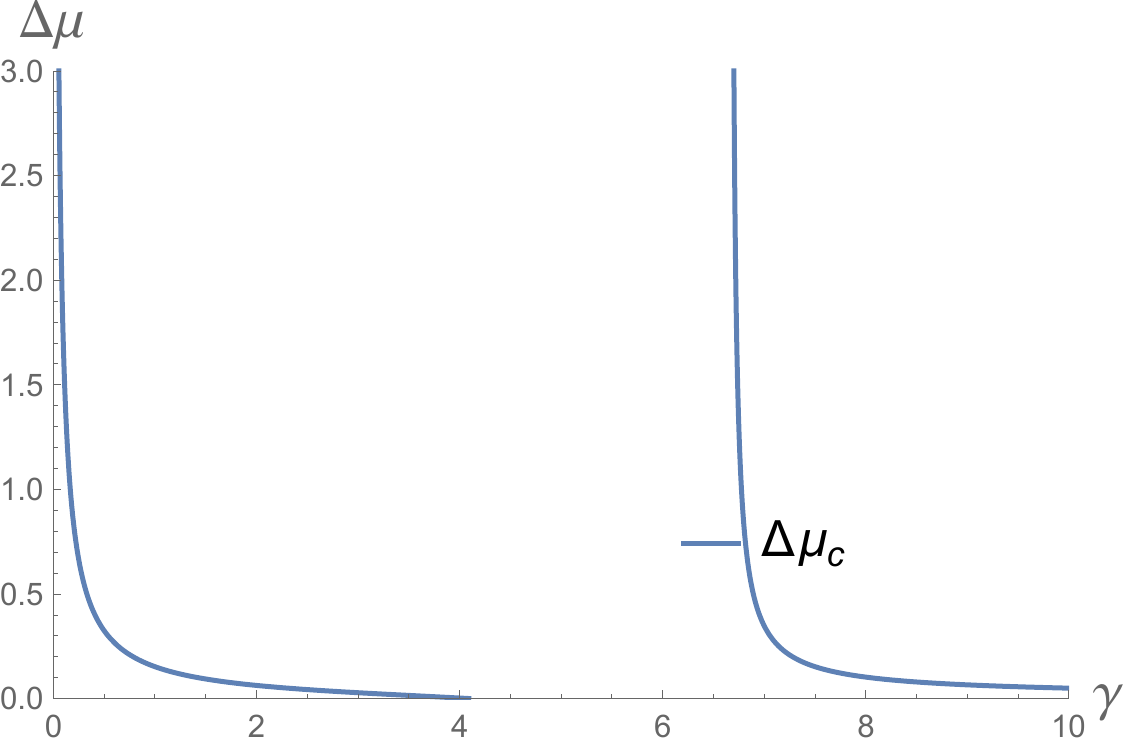}}~~~
	\subfloat[Parallel anchoring with contractile stress. $\zeta=-1,\nu=-0.27$]{\includegraphics[scale=0.25]{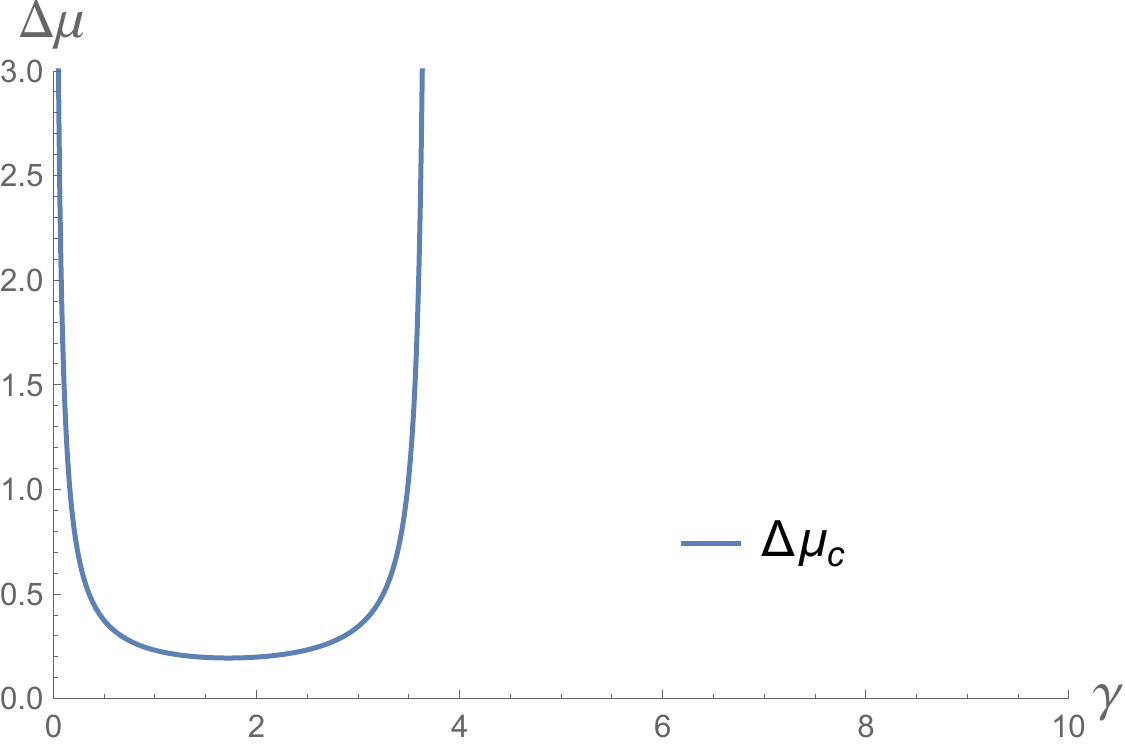}}~~~
	\subfloat[Perpendicular anchoring with extensile stress. $\zeta=1,\nu=0.4$]{\includegraphics[scale=0.25]{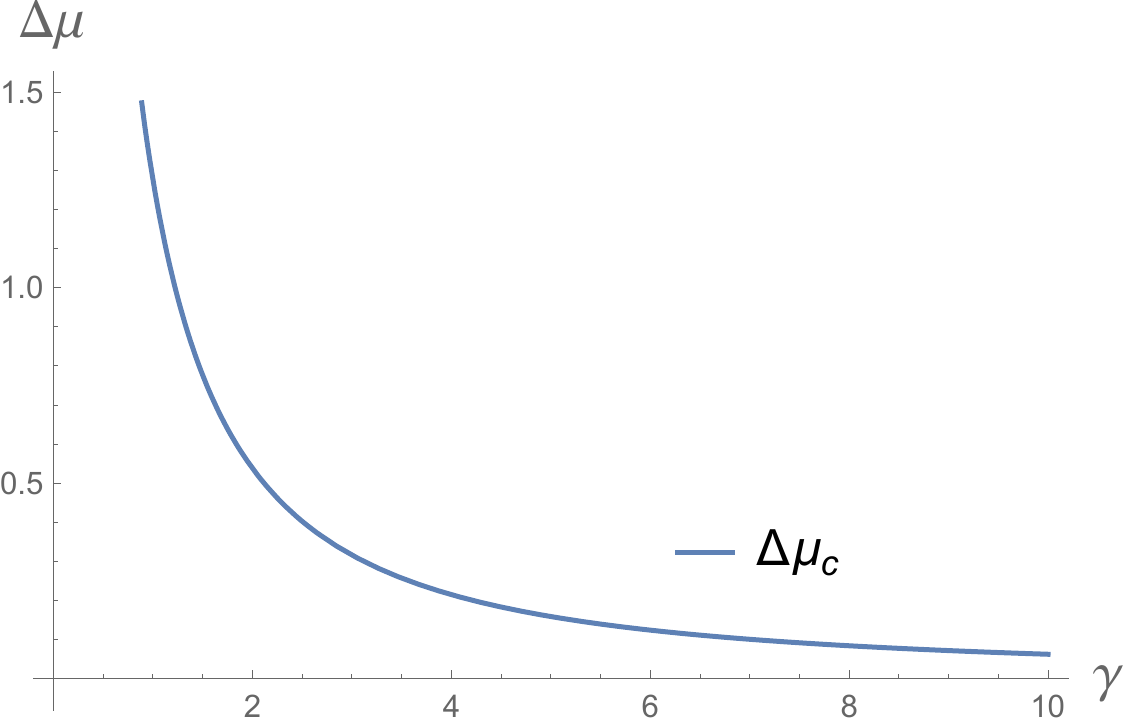}}
	\caption{Dependence of the critical activity $\Delta\mu$ on the rotational viscosity $\gamma$.}
	\label{fig:depgamma}
\end{figure}

Assuming $u_{xz}\neq0$, $u_{yz}\neq0$, but $u_{xy}=0$, thus allowing polarity to vary in both the X and Y directions, $(\theta(x,y),\phi(x,y))$, and using $\mathbf{h_\perp}=(0,-K\Laplace_{\{x,y\}}\kappa(x,y),0)$, we find the two-dimensional perturbation mode that corresponds to the out-of-plane wrinkling transition under extensile active stress as described in the main text and in the supplementary Mathematica notebook.
In this case, the strain rate up to linear order of tilts is
\begin{equation}
	\frac{\partial v_z(x,y)}{\partial x}=\frac{2 \Delta  \mu  (\gamma  \lambda  \nu +\zeta )}{4 \eta +\gamma  \left(\nu -1\right)^2}\kappa(x,y).
\end{equation} 
We study the effect of the rotational viscosity $\gamma$ on the critical activity in figure \ref{fig:depgamma}. Further behaviors can be predicted from the supplied Mathematica notebook applet.

